\documentclass[twocolumn,prl,aps,showpacs,superscriptaddress,floatfix,nofootinbib,10pt]{revtex4}

\usepackage{amsfonts,amsmath,amssymb}
\usepackage{graphicx}
\usepackage{color}
\usepackage[normalem]{ulem}
\usepackage{amsfonts}
\usepackage[toc,page]{appendix}
\usepackage[ocgcolorlinks,colorlinks=true,linkcolor=blue,citecolor=red]{hyperref}
\usepackage{latexsym}
\usepackage{amsfonts}
\usepackage{algpseudocode}
\usepackage{amsthm}
\usepackage{mathrsfs}
\usepackage{natbib}
\usepackage{color,verbatim}
\usepackage{psfrag}

\newcommand{\be}{\begin{eqnarray}}
\newcommand{\ee}{\end{eqnarray}}

\begin{document}
\title{Increased Certification of Semi-device Independent Random Numbers using Many Inputs and More Postprocessing}
\author{Piotr Mironowicz}
\affiliation{Department of Algorithms and System Modelling, Faculty of Electronics, Telecommunications and Informatics, Gda\'{n}sk University of Technology, Gda\'{n}sk 80-233, Poland}
\affiliation{National Quantum Information Centre in Gda\'{n}sk, Sopot 81-824, Poland}
\author{Armin Tavakoli}
\affiliation{National Quantum Information Centre in Gda\'{n}sk, Sopot 81-824, Poland}
\affiliation{Department of Physics, Stockholm University, S-10691, Stockholm, Sweden}
\author{Alley Hameedi}
\affiliation{Department of Physics, Stockholm University, S-10691, Stockholm, Sweden}
\author{Breno Marques}
\affiliation{Department of Physics, Stockholm University, S-10691, Stockholm, Sweden}
\affiliation{Instituto de F\'isica, Universidade de São Paulo, P.O. Box 6 6318, 05315-970 São Paulo, Brazil}
\author{Marcin Paw\l{}owski}
\affiliation{National Quantum Information Centre in Gda\'{n}sk, Sopot 81-824, Poland}
\affiliation{Institute of Theoretical Physics and Astrophysics, University of Gda\'{n}sk, 80-952 Gda\'{n}sk, Poland}
\author{Mohamed Bourennane}
\affiliation{Department of Physics, Stockholm University, S-10691, Stockholm, Sweden}


\date{\today}


\begin{abstract}
Quantum communication with systems of dimension larger than two provides advantages in information processing tasks. Examples include higher rates of key distribution and random number generation. The main disadvantage of using such multi-dimensional quantum systems is the increased complexity of the experimental setup. Here, we analyze a not-so-obvious problem: the relation between randomness certification and computational requirements of the postprocessing of experimental data. In particular, we consider semi-device independent randomness certification from an experiment using a four dimensional quantum system to violate the classical bound of a random access code. Using state-of-the-art techniques, a smaller quantum violation requires more computational power to demonstrate randomness, which at some point becomes impossible with today's computers although the randomness is (probably) still there. We show that by dedicating more input settings of the experiment to randomness certification, then by more computational postprocessing of the experimental data which corresponds to a quantum violation, one may increase the amount of certified randomness. Furthermore, we introduce a method that significantly lowers the computational complexity of randomness certification. Our results show how more randomness can be generated without altering the hardware and indicate a path for future semi-device independent protocols to follow.
\end{abstract}



\maketitle

Randomness is an important concept that manifests itself in many fields of science including statistics, biology, finance, informatics, social sciences and physics. Random numbers have vast applications in e.g. statistical sampling, Monte-Carlo simulations, cryptography and completely randomized designes. However, as John von Neumann aptly put it: "Any one who considers arithmetical methods of producing random digits is, of course, in a state of sin." Since knowledge of the program governing a software renders the output predictable, any such software is limited to produce  {\it pseudorandom} numbers. The use of pseudorandom numbers in tasks which require genuinly random numbers can lead to qualitative compromises in the task performance e.g. security breaches in cryptographic systems \cite{NSA,DG09}.

However, in quantum theory genuine randomness is a fundamental feature of the physical reality of quantum systems. Therefore, hardware based on quantum systems were proposed for generation of random numbers e.g. path-splitting of photons \cite{JA00}, the phase noise of a laser \cite{JC11, GTL10}, radio active decay \cite{YH99}, Raman scattering \cite{BD13}, and the arrival time of photons \cite{WL11}. Yet, how can we trust that the generated random numbers are not subject to some underlying predictability originating from the construction of the hardware, i.e. how can we be sure that the hardware is not just classically simulating the quantum system? To resolve this issue, the notion of device-independence \cite{MY98} was developed in which no assumptions are made on the inner workings of the hardware. By exploiting quantum correlations violating Bell's inequality \cite{Bell, CHSH} it was demonstrated that true random number generation is possible, even if we do not trust the supplier of our hardware \cite{PA10}. Unfortunately, device-independent protocols have strong requirements on their devices which leads to very low number generation rates, even with state-of-the-art technology. Semi-device independent (SDI) protocols  were proposed \cite{PB11} as good compromise between security and efficiency. In an SDI protocol, the devices remain untrusted but an upper bound on quantum channel capacity is assumed. This approach can be used for true random number generation \cite{LYW11,LPY12} and was also experimentally realized \cite{sdiexp1,sdiexp2}.

In an SDI protocol, we have no knowledge of the parameters (states and measurement operators). Therefore, to compute a lower bound on the amount of randomness generated, we need to optimize over all parameter settings that could reproduce the observed data, and then choose the least random result. Unfortunately, the target function is a quadratic function of the parameters and there are no known algorithms which guarantee to find a global minimum of such functions. This makes the optimization highly non-trivial. The most common approach is to use a semi-definite relaxation of the problem, i.e. optimize over a larger set such that it can be parameterized by variables in which the target function is linear. The first methods based on this idea were proposed in \cite{HWL13,HWL14}. Later they were replaced by a more efficient method from \cite{MiguelVertesi}. In this paper we investigate the computational requirements of these methods on postprocessing of experimental data for randomness certification. We show that there exists an interesting trade-off: the more computational power the user has to analyze the experimental data, the lower are the requirements on the experimental setup serving as hardware for randomness generation. Our results are both qualitative: a user with more computational power can certify the existence of randomness in a setup in which a user with less power cannot; and quantitative: given the same setup, a user with more computational power can certify more randomness. We also show how to reduce the computational complexity of randomness certification.

Our paper has the following structure. First, we describe SDI random number generation protocols. Then we discuss the methods used for randomness certification. Next we consider a particular quantum protocol, present its experimental realization, and apply our methods to analyze the experimental data. We conclude by a discussion of our results.

{\it Semi-device independent random number generation protocols.---} The structure of semi- or fully-device independent random number generation protocols is the same. The experiment is divided into rounds. Some rounds are chosen for security parameter estimation while the rest are used for generation of randomness.

We divide the total number of rounds into many groups of random size. In every group, the first round is used for parameter estimation and all remaining rounds in the group are used for randomness generation. Let $\mathcal{X}$ denote a set of all possible inputs the devices can have in a round and $\mathcal{X'}$ its arbitrary subset. For every round used for parameter estimation the inputs are randomly chosen from $\mathcal{X}$. An input is randomly taken from $\mathcal{X'}$ independently for every group, and then used in all randomness generation rounds within the respective group. In the standard device-independent random number generation protocols, $\mathcal{X'}$ consists of only one element i.e. the same inputs are used for all randomness generation rounds \cite{PA10}. Here we study a more general case. In fact, the number of possible settings of an randomness generation round (i.e. the number of elements in $\mathcal{X'}$), which we denote by $K$, is the parameter of the protocol that is responsible for both the computational complexity and the amount of randomness generated.\footnote{Examples of protocols with $K>1$ exist in literature, see e.g. \cite{M-delexp,moreRND} but they do not study the amount of randomness and the complexity of its certification as a function of $K$.}

The idea behind the class of protocols presented above is that if sufficiently many of the rounds are used for parameter estimation, e.g. $\mathcal{O}(\sqrt{N})$, then, because this set was chosen randomly, the average value of the parameter estimated for these rounds is close to the average for the remaining rounds (see \cite{PA10} for details). In fully device independent protocols this parameter is the violation of some Bell inequality, whereas in SDI protocols it is the efficiency of some communication game \cite{CCPs} in which the amount of communication is restricted in accordance with the known dimension of the quantum system which is assumed in SDI protocols. In both cases we know that values of these parameters imply lower bounds on the average amount of randomness in the measurement outcomes in each round. Our focus is the SDI approach in which one part of the device (Alice) receives the input $Z$ from which she prepares a quantum state $\rho_Z$ about which we only know the Hilbert space dimension $d$ (this is the SDI assumption). $\rho_Z$ is then sent to the other part of the device (Bob) who recieves his input $Y$ from which he determines  a measurement to perform on $\rho_Z$. The pair $X=(Z,Y)$ constitutes what we previously have called the input of the device in a given round. The result of Bob's measurement is denoted by $B$ and the whole procedure yields a corresponding conditional probability distribution $P(B=b|X=x)$. We use the following quantity as security parameter which allows us to estimate the randomness: $T=\sum_{b,x}c_{b,x}P(B=b|X=x)$. Now we describe the methods which we can use to certify randomness.

{\it Randomness certification.---} The randomness of the variable $B$ is quantified by conditional min-entropy, defined as
\be
H_{\infty}(B|X=x)=-\log \max_{b} P(B=b|X=x).
\ee
Our task is to find a lower bound on this quantity as a function of the parameter $T$. To this end we use methods from Ref. \cite{MiguelVertesi} based on semi-definite programming \cite{SDP} which are currently the state-of-the-art for this kind of problems. However, these methods are only able to optimize target functions which are linear in probabilities, which is not the case for $-\log \max \{\cdot\}$. Since $-\log(\cdot)$ is a strictly decreasing function, finding its minimum is equivalent to finding the maximum of the argument. The $\max$ part can be managed by performing a separate maximization for all $b$ and then choosing the largest value. This would be sufficient if the same setting was chosen for each round that is used for randomness generation. However, in the more general case we are interested in, we have to use the average min-entropy\footnote{We use this formula because, if it is later multiplied by the number of rounds, it represents log of the probability to guess the whole set of outcomes $B$ with the knowledge of the settings for each round available. This is the quantity in which we are usually interested when generating randomness.}
\be
H_{\infty}^{av}(B|X)=-\frac{1}{K}\sum_{x\in \mathcal{X'}}\log \max_{b} P(B=b|X=x).
\ee
Again, we can deal with $\log(\cdot)$ easily: using its concavity we have $H_{\infty}^{av}(B|X)\geq-\log\frac{1}{K}\sum_{x\in \mathcal{X'}} \max_{b} P(B=b|X=x)$ and thus we can focus on maximizing the argument. We should perform a separate maximization for every value of $b$, but this time we have to choose a separate value of $b$ for every element of the sum. This implies $D^K$ optimizations, where $D$ is the number of possible values of $B$. We see that the amount of computation grows exponentially with $K$ so we need a good reason for choosing $K>1$. Now, we will present a simple and intuitive reason for taking $K>1$, especially for protocols in which systems of high dimension are communicated. Later we show that our intuition is correct by considering a particular example.

Let's assume $K=1$ and consider a device which in Bob's part uses the optimal measurements for reaching the maximal value of $T$. The states for Alice are optimal for all $z$ apart from a particular one denoted $z_0$. Alice's state $\rho_{z_0}$ is an eigenvector of one of Bob's measurements, call it $y_0$. The values of these inputs are chosen in such a way that $x_0=(z_0,y_0)$ is the only member of $\mathcal{X'}$ i.e. only rounds with the input $x_0$ are used for randomness generation. Obviously, in this case there is no randomness as we can with certainty predict the measurement outcome. This comes at a price of lowering the value of the elements associated to $x_0$ in the sum in $T$. However, all the other elements still have the optimal quantum value and the overall change to $T$ is not significant. The impact of this is particulary strong for high dimensions since the possible values of $Z$ required for impossibility of achieving the maximal quantum value of $T$ with a classical protocol is greater than $d$ \cite{GBHA10}. The more values of $Z$ different from $z_0$, the less $T$ is decreased by the procedure described. It is easy to see why taking larger $K$ should help. In order to obtain no randomness for many different values of $X$, more elements of the sum in $T$ have to be below the optimal value.

For example, if the devices use the strategy described in the paragraph above, for the experiment that we describe in this paper, the critical value of $T$ below which no randomness can be generated is $T_{K}=\frac{16-K}{16}\frac{3}{4}+\frac{K}{16}\frac{5}{8}$. This value is obtained by noticing, that there are 16 possible inputs for Alice and if she sends the optimal state the success probability is $\frac{3}{4}$, while if she sends the eigenvector of one of Bob's measurements it is only $\frac{5}{8}$. For $K=1$ $T_1\approx 0.742$ and we see that the lower bound for larger $K$'s in fig.1 allows to certify the randomness for this value of $T$. The same behaviour is seen for $K=2$ and corresponding $T_2 \approx 0.734$. This clearly shows the advantage of using larger $K$'s. Let us now present this in more detail.

{\it The security parameter.---} The first SDI random number generation protocol \cite{LYW11} was based on a communication game in which Alice's input is two bits $z=(a_0,a_1)$ and Bob's input is a single bit $y$. Alice may communicate a two-level quantum system to Bob who aims to access the bit $a_y$ i.e.  the security parameter is $T=\frac{1}{8}\sum_{a_0,a_1,y}P(B=a_y|Z=(a_0,a_1),Y=y)$. This task is known as a quantum random access code \cite{ANTV99}.

The quantum random access code can be generalized to a multi-dimensional scenario: Alice's input numbers $a_0$ and $a_1$ can attain values from $0$ to $d-1$, and she communicates a $d$-level quantum system to Bob who aims to find $B=a_y$ \cite{THMB15}. In this work, we consider the particular instance of the multi-dimensional quantum random access code with $d=4$, the efficiency ($T$)  of which will serve as our security parameter;	
\be
T=\frac{1}{32}\sum_{a_0,a_1,y}P(B=a_y|Z=(a_0,a_1),Y=y).
\ee

\begin{figure}[t]
        \includegraphics[width=\columnwidth]{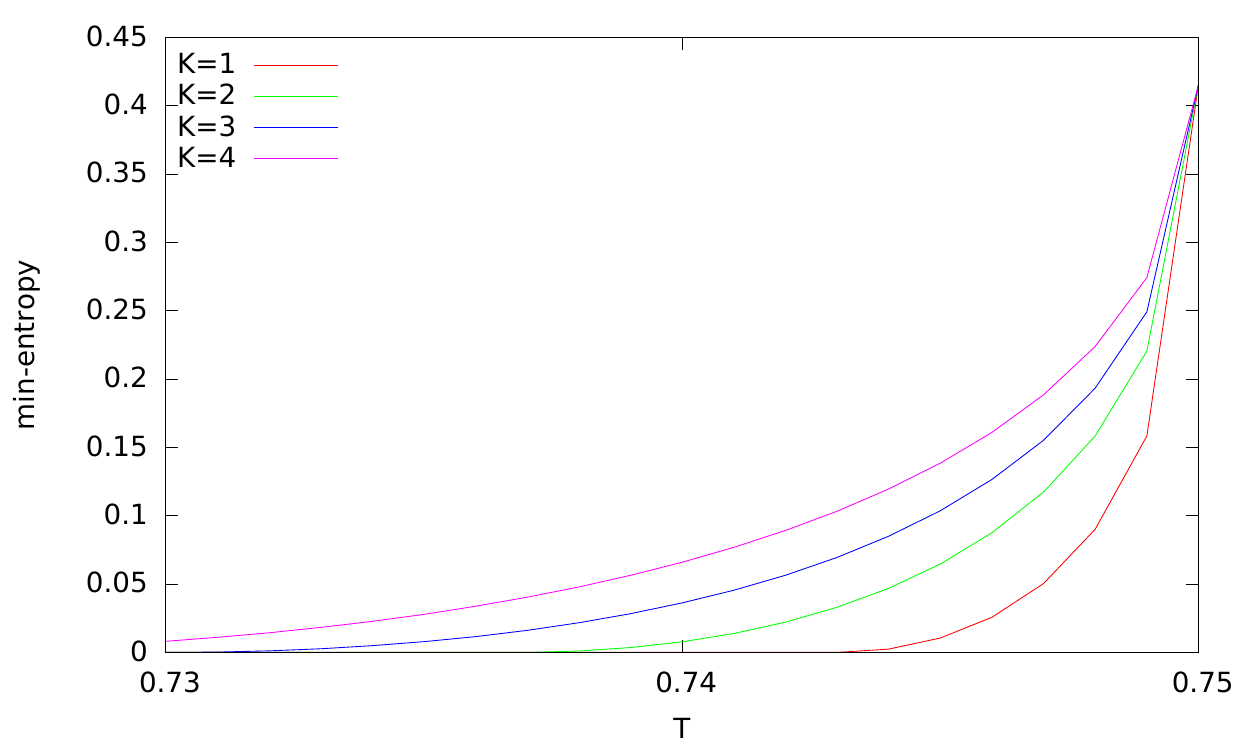}
        \caption{A lower bound on min-entropy is given for different values of $K$. Note that for maximal quantum value of $T$ the same amount of min-entropy is obtained. This amount is 0.4 which is much larger than 0.23 observed for a protocol based $d=2$ quantum random access code in \cite{LPY12}. This is one of the advantages of using quantum systems with a larger Hilbert space for communication \cite{THMB15}.    \label{fig:guessingPlot}}
\end{figure}

{\it Main results.---} We have applied the methods of Ref. \cite{MiguelVertesi} to evaluate the amount of randomness generated by the family of protocols based on the $d=4$ quantum random access code for different values of $K$. We have used a standard desktop computer on which a single optimization takes about 5 minutes. For arbitrary $K$, using standard methods, we would need $4^K$ optimizations for certification which quickly becomes impractical. For instance, $K=4$ would amount to roughly $21$ hours of computing for a single point on the highest line in fig.\ref{fig:guessingPlot}, if it was not the case that we came up with a method to reduce the computational complexity of the optimization. To achieve a reduction of computational complexity, we have exploited the properties of min-entropy and random access codes. The former depends only on the largest value of probability distribution while the latter effectively only distinguishes between $B=a_y$ and $B\neq a_y$.  Therefore, we have introduced a new binary variable $B'=if(B=a_y)$ which takes the value $0$ only when $B=a_y$, and 1 otherwise. Because $B'$ is obtained from $B$ by classical post-processing, the randomness of $B'$ is at most equal to that of $B$. Whether we do observe losses in entropy while moving from $B$ to $B'$  depends on the value of $T$. In the regime of large $T$'s, which is the one we are interested in, this will not happen. This is because the most probable value of $B$ is going to be the one for which the guess is successful (i.e. $B=a_y$ and $B'=0$) and min-entropy depends only on the highest probability in the distribution. If, for at least one of Alice's inputs, the most probable outcome of Bob would be different then the success probability would be lower than $T_{crit}=\frac{15}{16}\frac{3}{4}+\frac{1}{16}\frac{1}{4}\approx 0.72$. For larger values of $T$ we are sure that our method does not lead to the decrease of entropy. Our numerics suggest that the same happens for lower values.

The main advantage of using $B'$ instead of $B$ is that the former takes only two values and the number of optimizations needed to lower-bound the entropy is therefore $2^K$. Observe that this number of optimizations would remain unchanged even if we were to consider a quantum random access code of much higher dimension than $d=4$.

In fig.\ref{fig:guessingPlot} we have plotted the optimization results for different values of $K$ as a function of the security parameter $T$. We observe that for larger $K$, not only more randomness is certified but also the critical value of $T$, below which randomness is no longer certified, is lower.

To see how our analysis is relevant to the real experimental scenario we have performed an experimental realization of the quantum random access code with $d=4$ which we describe below.

{\it The experiment.---} We have implemented the security parameter estimation for a class of randomness generation protocols based on the $d=4$ quantum random access code studied in \cite{THMB15}.
The physical systems are defined by path and polarisation of single photons. The information is encoded in four basis states: $|1\rangle \equiv |H,a\rangle$, $|2\rangle \equiv |V,a\rangle$, $|3\rangle \equiv |H,b\rangle$ and $|0\rangle \equiv |V,b\rangle$, where   ($H$) and  ($V$) are horizontal and vertical polarization photonic modes respectively, and ($a$ and $b$) are two spatial modes of single photons. Any ququart state can be written as $a |H,A\rangle + b |V,A\rangle + c |H,B\rangle + d |H,B\rangle$. We have used a heralded, single photon source. The photons were generated through a spontaneous parametric down-conversion (SPDC) process where the idler photon is used as trigger. The emitted signal photon modes are coupled into a single mode fiber (SMF) and passed through both a narrowband interference filter (F) and a polarizer oriented to horizontal polarization direction.
Alice can produce any of the 16 states required by the protocol $|\psi_{a_0a_1}\rangle$ with $a_0,a_1\in \{0,1,2,3\}$ by  suitably oriented half-wave plates HWP($\theta_1$), HWP($\theta_2$) and HWP($\theta_3$), polarization beam splitter (PBS),  and  a setting of a phase shifter PS($\phi$).

\begin{figure}[t]
\includegraphics[width=\columnwidth]{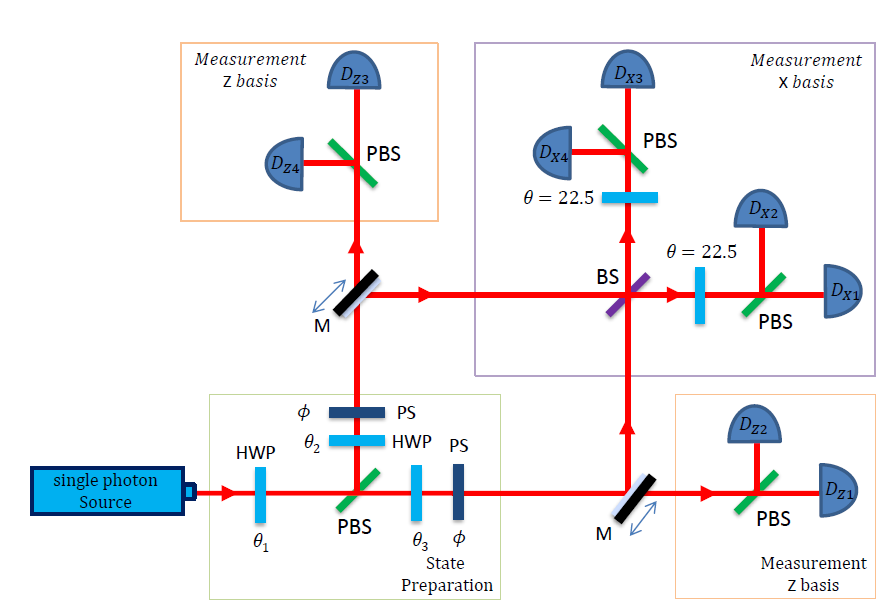}
\caption{Experimental set-up for the estimation of the security parameter $T$. Alice's quantum states are prepared through a combination of three suitably oriented half-wave plates, HWP($\theta_1$), HWP($\theta_2$) and HWP($\theta_3$), a polarization beam splitter (PBS) and a phase shifter PS($\phi$). Two mirrors M are used to realize Bob's choice of measurement basis. Detectors $D_{Zi}$ are associated to the $i$'th outcome of measurement $Z$ and similarly for $D_{Xi}$.}
\label{Fig2}
\end{figure}

Bob chooses between two measurement settings. The choice to measure in a particular basis is implemented by moving the mirrors ($M$) in and out with help of pico motor translation stages. For the computational basis ($Z$), the two removable mirrors are not present and the signal from detectors $D_{Zi}$ correspond to measurement outcome $i$. For the measurements in the Fourier basis ($X$) the mirrors are in place and the two spatial modes interfere at BS. In this case the measurement outcome $i$ corresponds to the signal from detector $D_{Xi}$.

Our single-photon detectors, both for trigger and  measurements, were silicon avalanche photodiodes with detection efficiency $\eta_d = 0.55$. All coincidence counts between the signal and idler photons were registered using a multi-channel coincidence logic with a time window of $1.7$~ns. The measurement time used for each experimental setting was $10$~s and the number of detected photons was approximately $2500$ per second.

The results we observed are in very good agreement with the predictions of quantum mechanics. We have observed $T=0.7347$, while the maximum that can be obtained with quantum resources is $T=0.75$.

{\it The results.---} We have used $T=0.7347$ for the estimation of randomness. The results for protocols with different $K$ are given in table \ref{tab:certRandomness}. $P_{av}(B'=0)$ is the average probability that $B'=0$ if the input is from $\mathcal{X'}$, i.e. $P_{av}(B'=0)=\frac{1}{K}\sum_{x\in \mathcal{X'}} P(B=a_y|(a_0,a_1,y)=x)$. $H_{\infty}^{av}$ is equal to $-\log P_{av}(B'=0)$. First we notice that for the standard protocol, with $K=1$ no randomness is generated despite the high fidelity of the experiment. However, the amount of randomness increases quickly with $K$. This comes at the price of  an increased number of optimizations. Nevertheless, it is a reasonable one time cost when we use the device for the first time because it is likely that later the same (or very similar) value of $T$ is going to be observed.

\begin{table}[t]
	\centering
		\begin{tabular}{|r|c|c|l|}
			\hline
			$K$ & $P_{av}(B'=0)$ & $H_{\infty}^{av}$ & time taken \\ \hline
			$1$ & $1$ & $0$ & c.a. 10 min.  \\
            $2$ & $1$ & $0$ & c.a. 20 min.\\
			$3$ & $0.99512$ & $0.007058$ & c.a. 40 min.  \\
			$4$ & $0.98180$ & $0.026499$ & c.a. 1.5h. \\
			$5$ & $0.96882$ & $0.045699$ & c.a. 2.5h. \\
			$6$ & $0.95565$ & $0.065446$ & c.a. 5h. \\
			$7$ & $0.94628$ & $0.079661$ & c.a. 11h. \\ \hline
		\end{tabular}
	\caption{Amount of randomness generated for different $K$ from the  experimentally obtained value of the security parameter using optimized method. In the last column the time it took us to perform the numerics is given, which reflects the increased complexity of larger $K$. It is the one-time cost that the user has to pay for before running the protocol. Later he just needs to check if the value of $T$ does not change. If it does the optimization has to be repeated for the new value. We estimate that to certify the randomness when all the settings are used for its generation, i.e. $\mathcal{X=X'}$ it would take 400.000 years on our machine.}
	\label{tab:certRandomness}
\end{table}

{\it Discussion.---}
We have presented a generalization of semi-device independent random number generation protocols to the case in which the randomness is extracted from more than one choice of inputs. We have shown that this approach can be used to certify more randomness without altering the experimental setup. This comes at a price of much higher requirements on classical computational power. Furthermore, we have shown how to significantly reduce the computational complexity of certification. We provided an intuitive explanation of origin of the advantages of our approach as well as demonstrated it in practice by performing an experiment and computing the randomness it generated. Nevertheless, we emphasize that there is no proof that another randomness certification algorithm that performs even better than ours does not exist. However, our results constitute a significant advance and indicate the direction which the research on quantum random number generation is likely to follow.

{\it Acknowledgments.---} This work is supported by Swedish Research Council (VR), ADOPT, ERC grant QOLAPS, FNP programme TEAM and NCN through grant 2014/14/E/ST2/00020. SDP was implemented in the free software package GNU OCTAVE\cite{octave} using the SeDuMi\cite{SeDuMi102,IntPoint} toolbox.

\end{document}